\documentclass[5p,times]{elsarticle}

\usepackage{lineno,hyperref}
\usepackage{acronym} 
\usepackage{adjustbox} 
\usepackage{array}
\usepackage{caption} 
\usepackage{subcaption} 
\usepackage{xcolor} 
\usepackage{listings} 
\journal{Entertainment Computing}

\makeatletter
\def\ps@pprintTitle{%
  \let\@oddhead\@empty
  \let\@evenhead\@empty
  \def\@oddfoot{\reset@font\hfil\thepage\hfil}
  \let\@evenfoot\@oddfoot
}
\makeatother

\begin{document}

\begin{frontmatter}

\title{Towards a low-cost universal access cloud framework to assess STEM students}

\author[comillas,iit]{L.F.S. Merchante\fnref{affcomillas,affiit}\corref{mycorrespondingauthor}}
\cortext[mycorrespondingauthor]{Corresponding author}
\ead[url]{https://web.comillas.edu/profesor/lfsanchez}
\ead{lfsanchez@comillas.edu}

\author[comillas]{Carlos M. Vallez\fnref{affcomillas}}
\ead{cmvallez@comillas.edu}

\author[comillas]{Carrie Szczerbik\fnref{afflang,affcomillas}}
\ead{cszczerbik@comillas.edu}

\fntext[affcomillas]{School of Engineering ICAI, Comillas Pontifical University}
\fntext[affiit]{Institute for Research in Technology, Comillas Pontifical University}
\fntext[afflang]{Modern Language Institute, Comillas Pontifical University}

\address[comillas]{25 Alberto Aguilera Street, 28015 Madrid, Spain}
\address[iit]{26 Santa Cruz de Marcenado Street, 28015 Madrid, Spain}

\begin{abstract}
Government-imposed lockdowns have challenged academic institutions to transition from traditional face-to-face education 
into hybrid or fully remote learning models. This transition has focused on the technological challenge to guarantee the
 continuity of sound pedagogy and to grant safe access to online digital university services. 
However, a key requisite involves adapting the evaluation process as well. In response to this need, the authors of this paper tailored 
and implemented a cloud deployment to provide universal access to online summative assessment of university students in a computer
 programming course that mirrored a traditional in-person monitored computer laboratory under strictly controlled exam conditions. 
This deployment proved easy to integrate with the university systems and many commercial proctoring tools. 
This cloud deployment is not only a solution for extraordinary situations, it can also be adapted to be used daily for online collaborative 
coding assignments, practical lab sessions, formative assessments, and masterclasses where the students connect using their own equipment.
 Connecting from home facilitates access to education for students with physical disabilities. It also allows participation and
 participating with their student’s own adapted equipment in the evaluation processes, simplifying assessment for those with 
hearing or visual impairments. In addition to these benefits and the evident commitment to the safety rules, this solution has proven
 to be cheaper and more flexible than on-premise equivalent installations.
\end{abstract}


\end{frontmatter}


\section{Introduction}\label{sec1}

Recent events have surprised most academic institutions. Very few could have imagined themselves in a forced and entirely online, sometimes asynchronous, environment. This transition has been challenging. Even the most innovative teachers who were used to introducing digital content in their lectures recognised that updating their method of instruction in just a few weeks´ time was extremely stressful. It was not just an issue of adapting material, training teachers in a set of tools or moving from a rigid timetable into a self-managed agenda. Infrastructures can also be a bottleneck. 

Most academic institutions, except those with generous funding and extremely large budgets, are not money-making machines. Thus, most did not have basic VPN (Virtual Private Network) technologies to provide their students with secure service to access university resources. And even if they had, the connection pattern to use those resources evolved as the students' transitioned from a face-to-face model into fully asynchronous remote learning. Infrastructures are no longer accessed massively in the same time slots.  On the contrary, the access to those resources is distributed throughout the day. This change of paradigm has severe implications for the IT (Information Technology) department, their services and administration. 

The impact of remote experimentation on laboratory learning due to COVID19 restrictions has already been highlighted \cite{achuthan2021impact}. Still, less attention has been given to how the assessments of the students have been impacted. Some authors have studied digital assessments \cite{begnum2018digital, bryan2019innovative} but few have proposed solutions to preserve exam conditions when students are at home \cite{gamage2020online, mahajan2021blended}. On the other hand, from the point of view of the effects on the mental health of the students, academic institutions closure have increased their levels of stress and anxiety \cite{jones2020student, wang2020impact, husky2020stress, basheti2021prevalence, wu2021increases, coakley2021anxiety} although this behavior has not been observed in all the countries \cite{johansson2021depression}. Several authors have revealed the negative relationship between the students anxiety and their academic results \cite{carver1988control,seipp1991anxiety,cassady2002cognitive,vickers2007performing,vitasari2010relationship,dordinejad2011relationship, owens2012anxiety,su2017does}. This effect has long been known to apply with subjects like computer programming \cite{chang2005computer}.
Evaluating practical skills, like computer programming, by the conventional ways (all sort of face to face evaluation on computer labs) are now neither possible nor practical. These limitations could have led to a detrimental effect on low-stake assessment in favour of high-stake. A form of retrieval practice, also known as the testing effect (students recalling the class material to be learned from their memories), this low-stakes testing helps to reduce such levels of stress and has extremely positive effects on student achievement \cite{von2018test}. Much more effective than rereading or restudying class materials in high-stakes evaluative contexts, active recall by actively reengaging with the material is the best strategy \cite{adesope2017rethinking}.

One of the main difficulties that academic institutions have come across has been evaluating their students. Many online programs require students to attend an on-site evaluation. Commuting to the exam is especially painful for mobility-impaired students, and using standard computers when the assessment involves the evaluation of digital competencies may be a severe disadvantage for hearing or visually impaired students that need special equipment. Convening students in person is not a mere whim; the reason is the validation of their identities, the inviolability of the exams, and the conditions of the environment are not straightforward when the students take the exams from home. Nevertheless, today there are a plentiful number of techniques that allow educators to administer exams safely without supervision. Creating random and timed tests \cite{howlett2005securing} is one of the most popular. There is a vast literature about evaluating digital competences \cite{sillat2021digital} with most of it based on different types of interviews, questionnaires and surveys. Unfortunately, these techniques are hard to apply to subjects where the purpose is to assess the students' skills with, for example, a programming language, which is the area of knowledge utilized for this paper. 

Although resources are available to solve some of the issues related to online assessment, academic institutions are not generally large enough to have an IT department trained on Big Data, cybersecurity or cloud services. These technologies are, however, becoming more and more common in academic environments. Big Data solutions were first introduced as a solution to a large amount of information and the scaling limitations of the traditional platforms. Academic institutions and other stakeholders from the educational system are facing challenges to deal with and profit from that amount of data. Over the last decade, there has been an increasing amount of literature in this area \cite{west2012big, reyes2015skinny, anshari2016developing, wang2016big, song2017, dahdouh2018big, williamson2018hidden,cantabella2019,miah2020editorial, williamson2020big, baig2020big, lnenicka2020big, tlili2021towards}.

With the widespread use of high-speed Internet connections worldwide, cybersecurity issues are now central to discussions in board rooms in every sector, and academic institutions are no exception. Two important tenets of higher education are academic freedom and openness \cite{ulven2021systematic} but also added as a third is the massive aggregate of computing power \cite{singar2020role}. This is the reason universities and research centers have become lucrative targets for cyber-attacks \cite{ncube2010lessons,bongiovanni2019least}. They handle sensitive personal data and valuable research results. Important companies in technology and media have published reports about the relevancy of cyber-threats to the educational sector \cite{dell2018, alcatelucent2020, sophos2021, redmond2021, security2021, osterman2021}; some of which apply exclusively to online learning \cite{saxena2012cyber, davidson2014cyber}. 

However, cloud services are underutilized in the education sector though they have been used in a limited manner to build cloud supported collaborative environments \cite{chao2015, vickers2015media, wang2016acceptance, al2018systematic, mosci2018, baanqud2020engagement}, e-learning platforms \cite{pocatilu2009using, bora2013learning, riahi2015learning, khan2020cloud, tsai2020investigating} or cloud storage \cite{stantchev2014learning, bond2018digital}. These are just three examples of many possible use cases that cloud-based tools could provide to higher education.

The purpose of this article is to share with the academic community the design and application of the online assessment of university students from STEM (Science, Technology, Engineering, and Mathematics) degrees remotely through a cloud deployment by simulating the environment in a monitored computer laboratory. This cloud infrastructure allowed for the adherence to the following actions: 

\begin{itemize}
	\item Verify the identity of the students 
	\item Secure their equipment to ensure that students only use approved resources
	\item Capture snapshots of the students' progress regularly. Not only for backup purposes but also to be able to reproduce the student timeline under a situation of fraud investigation
	\item Provide the student with secure access to the tools required for the subject evaluation with minimum intervention on their part
	\item Allow visually and hearing impaired students to use their own equipment
\end{itemize}

The infrastructure required for this project was deployed through cloud providers. Since many academic institutions do not have sufficient IT staff or knowledge to install and administer new platforms, cloud providers allow these deployments to be done quickly and to provide the best cost-effective utilization of resources. In addition, cloud deployments have several valuable strengths highlighted below:

\begin{itemize}
	\item They are easy to integrate with the university authentication servers to benefit from a single-sing-on system
	\item They can be designed to scale their resources automatically as a function of the computation requirements
	\item They can be adapted to many commercial proctoring tools
	\item They can be easily administrated with scripts, thus requiring no additional IT staff
	\item Once the exam is finished, data can be backed up and resources released, enabling the budget to remain under control when the platform is not needed
\end{itemize}  

\section{Material and methods}

The use case described in this document is providing a safe and flexible coding framework for STEM students.  Although useful for many academic applications such as practice, study and feedback, when used for exams during assessment and evaluation, a proctoring tool is required. Commercial proctoring tools are software products conceived to convert a personal computer into a trusted device. These tools limit usage by controlling the URLs (Uniform Resource Locator) that the students can browse or the applications they can execute. Most proctoring tools use webcams and microphones to record the environment and some locally installed software to restrict the applications that the student may attempt to open. As this is a relatively new market, there is relevant research emerging, such as facial recognition to detect cheating probability based on the students' faces or eye movement \cite{hylton2016utilizing}. There is an endless collection of proctoring tools, and selecting one that fits a particular set of requirements can be challenging \cite{foster2013, nigam2021systematic}. Any given academic institution generally chooses one to use in its facilities.

Most of these tools allow the introduction of URL exceptions to access their intranet or specific websites to access the resources required for the exams. However, additional software modules are necessary when assessing programming skills, such as compilers, language interpreters or editors. Integrating a coding framework with a proctoring tool would require wrapping the coding environment on a cloud website that could be accessed through a URL. This URL could be added to the proctoring tool list of exceptions.  This setup would benefit from the surveillance provided by proctoring tools and the flexibility of a cloud coding framework.

\subsection{The proctoring application}

The proctoring application used in this case study was \emph{Respondus} \cite{respondus2021}. This tool relies on installing an application called \emph{LockDown Browser}, a restricted version of a regular web browser that allows both limiting the URLs that the students can visit and introducing exceptions. At the same time, \emph{Respondus} locks the student's computer desktop, halting concurrent applications like email or messaging clients that would allow for cheating. This software uses students' webcams and microphones to guarantee the required test-taking conditions and applies artificial intelligence video processing algorithms to detect suspicious movements. This tool can be integrated with academic platforms like \emph{Moodle} \cite{moodle2021} to be the entry point for exams.

\subsection{The coding framework}

Since its appearance in 2001, \emph{iPython}, an interactive server-based coding environment \cite{ipython2007}, has gained in popularity. Code developed with \emph{iPython} resulted in \emph{notebooks} that were, and still are, HTML (HyperText Markup Language) documents where markdown cells coexist with Python code.  The evolution of this Python package, now renamed \emph{Jupyter} \cite{jupyter2021}, accepts a great variety of \emph{kernels}, providing its notebooks with the capacity to code in different languages \cite{kernels2021}. Its popularity has increased so much so that similar projects inspired by the \emph{Jupyter} project appeared with remarkable relevance. One example of these copycats is \emph{Google Colab} notebooks \cite{colab2021}. Its notebooks are like \emph{Jupyter's} with the added advantage that Google provides restricted access to GPUs without fees. Of course, development environments based on notebooks are far from replacing traditional IDEs (Integrated Development Environment) like \emph{Eclipse}, \emph{Pycharm} or \emph{Xcode} \cite{eclipse2021, pycharm2021, xcode2021}. 

Coding with IDEs becomes essential when the size of the project exceeds a minimum complexity. Yet, in fact, a significant percentage of developers cover their needs with notebooks with the advantage of its simplicity. In addition, those notebook-based environments like \emph{Jupyter}, now renamed to \emph{Jupyter Hub}, are evolving in exciting directions like the distribution \emph{Zero to Jupyter Hub} \cite{zero2021} that allows deploying a \emph{Jupyter Hub} over \emph{Kubernetes} \cite{kubernetes2021}.

\begin{figure}
\centering
\begin{subfigure}{.5\textwidth}
  \centering
  \includegraphics[width=.75\linewidth]{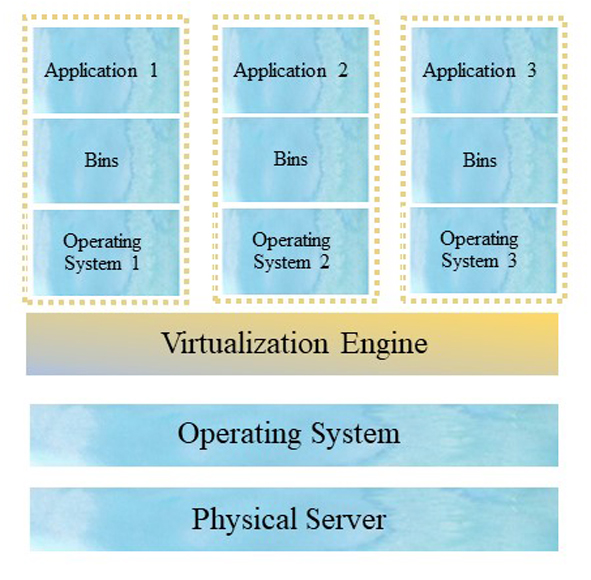}
  \caption{Classic Virtualization Stack Diagram}
\end{subfigure}
\vspace{0.5cm}
\begin{subfigure}{.5\textwidth}
  \centering
  \includegraphics[width=.75\linewidth]{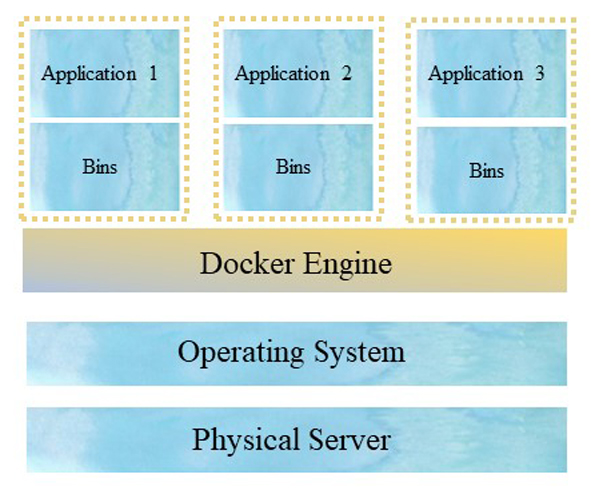}
  \caption{Docker Stack Diagram}
\end{subfigure}%
\caption{Classical Vs. Docker Virtualization}
\label{fig.docker.stack}
\end{figure}

\emph{Kubernetes} is an open-source container-orchestration system for automating application deployment, scaling and managing  \cite{wikikube2021}. A brief explanation about containers is required before understanding the usefulness of \emph{Kubernetes}. Acknowledging some loss of accuracy of the terminology on behalf of simplicity, working with containers can be described as a technique to deploy applications on a computer without passing through complex installations and configurations. Its most representative application is \emph{Docker} \cite{docker2021,anderson2015docker}. \emph{Docker} is a product that uses Operative System Virtualisation to deliver software packages in the shape of containers \cite{wikidocker2021}. Figure \ref{fig.docker.stack} shows the difference between classic virtualization and \emph{Docker's} containers.

\emph{Kubernetes} and \emph{Docker} are not new \cite{turnbull2014docker, bernstein2014containers}. The initial release of \emph{Kubernetes} appeared in 2014 after the initial release of \emph{Docker} in 2013 that was inspired by Linux containers created in 2008. Many authors saw the potential of those technologies that today have taken over the industry \cite{bernstein2014containers, brewer2015kubernetes, burns2019kubernetes}. The distribution \emph{Zero to Jupyter Hub} uses those two technologies to deploy as many instances of \emph{Jupyter} on a \emph{Kubernetes} cluster as needed by the real-time demand of students. As the audience for this article is not obliged to have deep knowledge of technology, the proposed solution is described below.

\begin{figure}[!ht]
\centering
\includegraphics[scale=1.1]{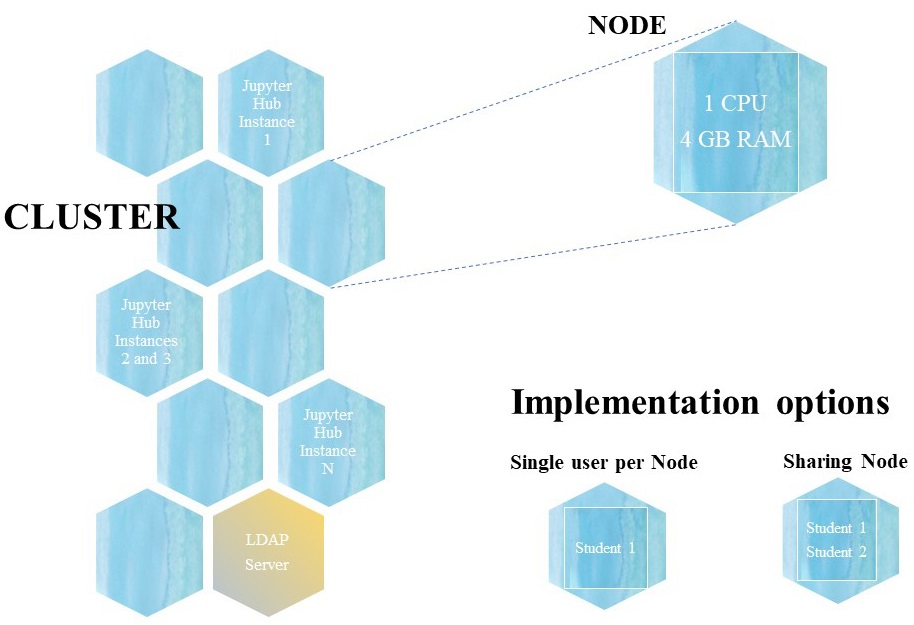}
\caption{Kubernetes Cluster High-level Diagram}
\label{fig.cluster.scheme}
\end{figure}

The proposed platform is supported by a \emph{Kubernetes} cluster formed by an ensemble of nodes running a piece of software (called \emph{middleware}) that allows the nodes to understand each other. Applications can be deployed on top of a \emph{Kubernetes} cluster, similar to a user installing an application on top of an operating system in a laptop. The advantage of deploying an application over \emph{Kubernetes} instead of using a regular computer is that the cluster contains far more resources than a PC; therefore, if the application requires more computational power (RAM, CPU…), the \emph{Kubernetes} cluster can start new application instances. The application to be deployed on top of \emph{Kubernetes} cluster in this particular case study in assessment is \emph{Jupyter}. Every time a student logs into the service, \emph{Kubernetes} creates a \emph{Pod}. A \emph{Pod} is an instance of \emph{Jupyter} with some limited resources reserved for that student. If the number of students increases and the resources start to become scarce, the deployment of \emph{Kubernetes} scales up automatically, adding more nodes to the cluster with more CPUs and more RAM to create more \emph{Pods}. Likewise, if the number of students decreases and some resources are released, \emph{Kubernetes} can scale down to save on the budget. No human intervention is needed in this task. Students need to worry only about their coding and teachers about ancillary tasks such as answering students' questions or capturing backups.
Figure \ref{fig.cluster.scheme} displays a simple diagram with general-purpose nodes (1 CPU and 4GB of RAM) showing that each node can host one or multiple instances of \emph{Jupyter} to service one or multiple students. The Figure \ref{fig.cluster.scheme} also includes the authentication service to be explained in a later section.

Installing and configuring such an infrastructure as described in the previous paragraphs is complex and not feasible for everyone. Cloud providers resolve this knowledge gap with their simplified interfaces, allowing a regular user to create a Kubernetes cluster by just filling out a form, clicking a button and waiting five minutes \cite{kubequickstart2021}. More details about deploying a similar infrastructure are given in \ref{replication}\footnote{The deployment is fully documented in repository \url{https://github.com/lmerchante/LowCostUniversalAccess}}.

\section{Deployment Considerations}
\subsection{The technological trade-off}
There is currently an animated discussion regarding the benefits and drawbacks of traditional computing versus cloud computing. On classic infrastructures involving traditional computing, also called \emph{on-premise}, the customer installs the computation nodes on-site and must deal with the operating system and other software installation and configuration, network creation and the platform's administration. On the other hand, cloud deployments are built using remote virtual machines that the provider boots in different geographic locations. There is a great deal of literature comparing both types of deployments \cite{bibi2012business, boillat2013premise}. The Internet has plenty of articles and use cases justifying the benefits of using one or the other. Still, most of the conclusions refer to a mere economic analysis. Unfortunately, the decision depends strongly on the use case and the size and background of the institution that is considering the deployment. On-premise infrastructures involve many expenses, especially for non-IT institutions; the following simplification provided here would include the following:

\begin{itemize}
	\item Physical location. Big storage platforms and computation engines must be installed in specific places with particular features. The nodes are extremely heavy, and some deployments involve racks with dozens of computation engines; therefore, not all floors are prepared to support such weight. In addition, CPUs and power sources emit very high heat, thus the cooling conditions of this space are a critical design feature. Finally, these locations must be chosen far from workplaces or lecture buildings because hundreds of heat-sinks fans and climatisation equipment produce high noise levels. 
	\item Power consumption. The energy required to power a computation platform is a significant component of a budget consideration. Each computation node can consume the same electricity as tens of standard laptops. In addition, depending on the number of resources requested, some computation platforms may need several nodes installed per rack, and some institutions may need several racks to service a large number of students. Finally, the consumption of the air conditioning system needs to be included in the institution's budget.
	\item Disaster recovery. This variable needs to be taken into consideration depending on the criticality of the system deployed. A platform as the one described in this document that would replace traditional computer labs needs to be failure resistant. This would mean being resilient in power outages as well as recovery from hardware failure with no data lost, primarily if used for taking exams. Disaster recovery can be granted by proper use of redundancy on the hardware and power systems.
	\item Human resources. These infrastructures must be monitored 24 hours a day. A power loss may shut down the whole platform impacting the connected users. A failure of the air conditioning equipment may increase the temperature of the data centre resulting in hardware failure. Thus, staff in charge of monitoring and administering the platform and a maintenance team are a requirement. In addition, depending on the data, security companies may be needed to grant the inviolability of the compound. Data integrity supports the activity of the companies, and suffering acts of vandalism in the data centre or cybernetical intrusions to the data might have serious consequences.
\end{itemize}

These four aforementioned dimensions have evident economic repercussions should an institution employ on-premises infrastructure. On the other hand, setting up a cloud deployment requires a laptop and a credit card. Cloud resources are billed by the number of hours of actual use, and their prices per hour are generally affordable. However, when the number of activity hours increases, costs become comparable.

Cloud deployments also have advantages regarding the software life-cycle. Software products, especially the most recent ones, evolve very fast. Many products undergo major upgrades at least once per year. Upgrading on-premise infrastructures involves more effort than upgrading in the cloud. Ignoring  updates would risk the institution's platform and result in the students using outdated products during their course of study or assessments. Due to the virtual nature of cloud deployments, most updates are not disruptive because newly updated nodes can be added to the deployment before removing the old ones without stopping the service. Non-disruptive updates can also be accomplished on-premise, but the technical, administrative and human complexity to schedule a major upgrade makes many companies ignore or postpone them sometimes indefinitely. 

Given the great number of variables involved in the trade-off \emph{on-premise} vs cloud, providing a basic rule of thumb would be inadequate. However, most cloud architects would agree on two basic approaches:

\begin{itemize}
	\item If the institution has a small IT department (less than ten people), the cloud is considered the best choice
	\item If the institution has the location, the background and the staff to establish something on-premise, then the decision will depend on the up-time. For example, if the infrastructure has to be running 24 hours, \emph{on-premise} would be the better suited option. On the other hand, if the use case can be limited to very few time slots, then cloud deployments would most likely be less costly.
\end{itemize}

Based on these two guiding recommendations and considering that many academic institutions do not have large IT departments, choosing cloud deployment might prove to be a means for efficient use of resources.

\subsection{The efficiency boost}

Using cloud applications appears simplistic on the surface. Providers have created very friendly portals to find whatever resources through different paths. At the same time, they deliver dashboards where at a glance, a user can review which resources are being used and the evolution of billing in real-time. Everything seems accessible, but this is a slightly biased first impression. 

Creating a \emph{Kubernetes} cluster, such as the one described earlier in the text, can be accomplished with a few clicks, selecting the number and the type of nodes. The whole deployment is completed in five minutes. Once the infrastructure is running, software needs to be installed, for example, \emph{Jupyter Hub}. This task may take some additional minutes. \emph{Jupyter}, like any other piece of software, requires proper configuration. A basic example might be setting up encryption for HTTP (Hypertext Transfer Protocol) communications and linking the cluster with an authentication server in order that the students can use their academic login service to sign into the programming framework.  Additional configuration is required to access the coding environment through a URL with a domain name instead of an IP address.

For those new to cloud services, the entire process of setting up an online, secured, auto-scalable programming environment could take three or four hours with successive setups taking a third of the time. The main drawback of deploying cloud services with the GUI (Graphical User Interface) is that it will take one hour to set up all the modules every time they are needed. One must bear in mind that cloud deployments are billed by use. If the purpose of a cloud programming framework, like this, is servicing exams, then it is only required to keep it running for two or three hours. Thus, after backing up the students' work, the resources used in the deployment will be released. This process needs to be reproduced for every use of the framework, which translates into one hour to boot and configure it properly and a few additional minutes to release the resources. Wasting one hour each time that the service is needed could be unacceptable.

One of the main strengths of cloud services is that any task can be completed using their APIs (Application Programming Interface). These APIs allow script creation to automate the deployment and release of any resource. Graphical interfaces are friendly, but scripts are faster and can be easily automated. The design of scripts requires some expertise, but the time invested in their development will be advantageous. Scripting the creation of the service may reduce the deployment time two or three times and, most importantly, without human intervention. Any script can be empowered with input arguments, such as the number of nodes or type of hardware, depending on whether the purpose of the deployment requires more memory or more CPU.

Basic scripting is very easy. For instance, a 20-node \emph{Kubernetes} cluster with auto-repair, auto-upgrade and auto-scaling activated can be created with the script shown in Script \ref{lst:script.create}. The type of node is \emph{n1-standard-4} that are general-purpose nodes with 4 CPUs and 15GB of RAM. Note that the creation script is enabling auto-scaling. If this option is disabled, then the cluster can be manually scaled up and down using Script \ref{lst:script.scaleup} and Script \ref{lst:script.scaledown}. When the exam is finished, the cluster can be removed with Script \ref{lst:script.released}. Those scripts are tiny because only a few parameters and environment variables were used on behalf of simplicity. Google Cloud Platform documentation shows more than fifty parameters to create the cluster most suitable to every need. Actual scripts can be found in the associated repository \footnote{Repository: \url{https://github.com/lmerchante/LowCostUniversalAccess}}.

\renewcommand{\lstlistingname}{Script}

\begin{lstlisting}[basicstyle=\footnotesize ,label={lst:script.create}, language=command.com,caption={Cluster Creation Script}, captionpos=b]]
ECHO OFF
CLS
SET PATH=%GOOGLE_CLOUD_SDK%\bin;%PATH%;
cd C:\Users\admin\AppData\Local\Google\Cloud SDK
ECHO Create Kubernetes Cluster
set REGION=us-central1
gcloud container clusters create icai-jupyter ^
  --region %REGION%^
  --num-nodes=20^
  --machine-type=n1-standard-4^
  --enable-autorepair^
  --enable-autoupgrade^
  --enable-autoscaling^
  --max-nodes=60^
  --min-nodes=10
ECHO ---
ECHO ON
\end{lstlisting}

\begin{lstlisting}[basicstyle=\footnotesize ,label={lst:script.scaleup}, language=command.com,caption={Cluster Scale Up Script}, captionpos=b]]
ECHO OFF
CLS
SET PATH=%GOOGLE_CLOUD_SDK%\bin;%PATH%;
cd C:\Users\admin\AppData\Local\Google\Cloud SDK
ECHO Add nodes cluster Kubernetes
gcloud container clusters resize icai-jupyter ^
  --node-pool default-pool --num-nodes 60 --quiet
ECHO Done
ECHO ON
\end{lstlisting}

\renewcommand{\lstlistingname}{Script}
\begin{lstlisting}[basicstyle=\footnotesize ,label={lst:script.scaledown}, language=command.com,caption={Cluster Scale Down Script}, captionpos=b]]
ECHO OFF
CLS
SET PATH=%GOOGLE_CLOUD_SDK%\bin;%PATH%;
cd C:\Users\admin\AppData\Local\Google\Cloud SDK
ECHO Remove nodes cluster Kubernetes
gcloud container clusters resize icai-jupyter ^
   --node-pool default-pool --num-nodes 10 --quiet
ECHO Done
ECHO ON
\end{lstlisting}

\begin{lstlisting}[basicstyle=\footnotesize ,label={lst:script.released}, language=command.com,caption={Cluster Release Script}, captionpos=b]]
ECHO OFF
CLS
SET PATH=%GOOGLE_CLOUD_SDK%\bin;%PATH%;
cd C:\Users\admin\AppData\Local\Google\Cloud SDK
ECHO Delete Kubernetes Cluster
set REGION=us-central1
gcloud container clusters delete icai-jupyter
ECHO ---
ECHO ON

\end{lstlisting}

Scripting actions allow scheduling deployments unattended on demand and automated resources release when there are no longer needed. This is the preferred scenario that provides the best utilization of resources versus budget trade-off and is the most appropriate for many academic institutions. 

\subsection{Resources are virtual, but money is real}

There exist studies about the benefices of cloud services for academic institutions \cite{koch2021}. However, tracking the consumption of cloud services can be tricky. Most cloud providers offer free credits to access their platforms and start deploying services. The experience of those who tried free tiers is that the credits vanish faster than expected. All cloud providers' portals have made great efforts to provide the users with visualisations and alerts in order to permit users to track how much money every GB of memory or CPU costs. Monitoring the billing is trivial; the problem is that including new cloud resources into the deployment is very easy. Increasing the storage, assigning permanent public IPs, adding an encryption service, requesting a domain name or adding extra nodes to the cluster are everyday tasks. Administrating the budget can be challenging when the number of resources becomes large. Cloud services can result in a higher initial cost than was initially estimated. However, this tends to improve with greater expertise and the use of automation to maintain the resources during the minimum time required.

\section{Results and Discussion}

The content presented in this article corresponds with an actual project carried out in the authors´academic institution of higher education to take remote exams to evaluate coding skills. The chosen subject was \emph{Introduction to Computer Science} which teaches the foundation of coding using Python. Due to the recent government-mandated health restrictions in schools, the evaluation was done remotely with the students at their homes, using proctoring tools to reproduce traditional exam conditions. 

\subsection{The exam}
The exam consisted of two parts, the first of which was a quiz with random, timed questions. Most academic platforms allow configurations of this kind of test on a \emph{Moodle} platform. The second section of the exam consisted of practical exercises where the students needed to write pieces of code. For this part, \emph{Jupyter Hub} was set in place in a cloud deployment in order that students could use a fully functional coding framework.

\subsection{The alternatives}
Two setups for online assessment had already been tested by the university on previous occasions:

\begin{itemize}
	\item Setup A: Students are split into small groups (six to twelve people) monitored by one teacher using a video-conferencing tool (\emph{Teams}, \emph{Zoom}, \emph{Webex}) and requiring the students to have their cameras and microphones enabled, implementing a remote version of the physical surveillance.
	\item Setup B: Assist the students in configuring the camera from their cell phones to capture the environment and the webcam from their laptops to capture their bodies. At the same time, another piece of software captures their screens. The video resulting from the recording software and both cameras must be combined on a single video stream and sent back to the teachers at the end of the exam. This mechanism required extraordinary configuration efforts the previous days and a reliable Internet connection to upload several MB of video.
\end{itemize}

Neither setup A nor setup B allowed for the use of proctoring tools because video-conferencing software is usually blocked when any of those tools are installed.

The setup proposed in this article, Setup C, consisted of an online coding environment, \emph{Jupyter Hub}, accessed through HTTPS integrated with a proctoring tool. Students verified their identity through an LDAP (Lightweight Directory Access Protocol) server, and backups of the students' exams were collected periodically using scripts.

\begin{figure}[!ht]
\centering
\includegraphics[scale=0.95]{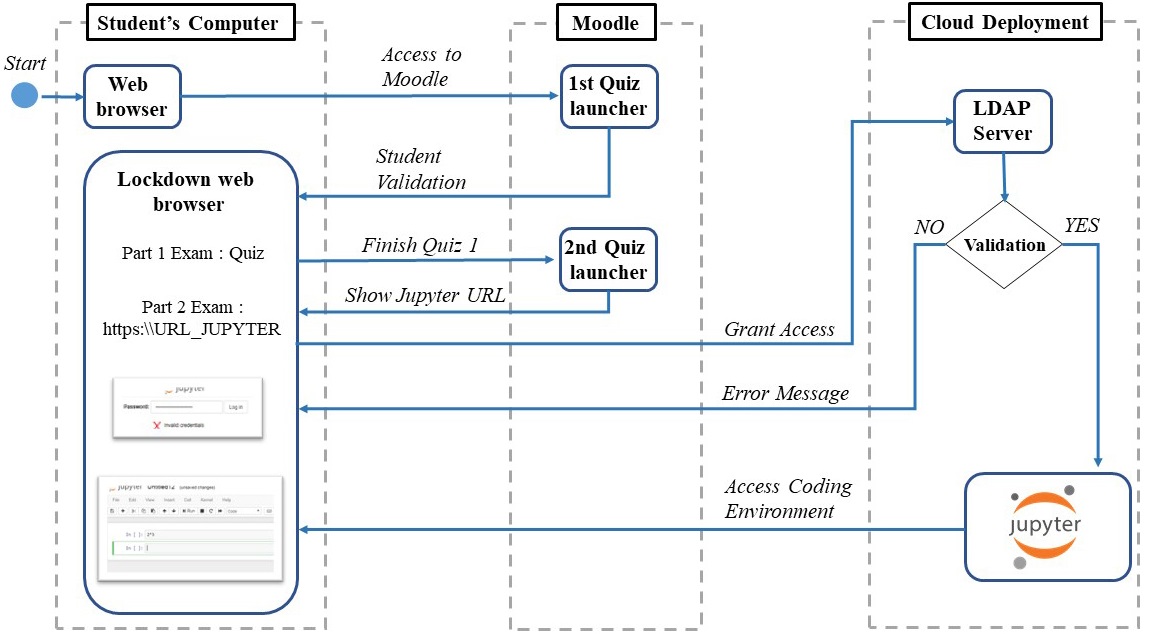}
\caption{Diagram of the Proposed Test}
\label{fig.exam}
\end{figure}

This setup allowed the integration of \emph{Jupyter Hub} with the proctoring tool, which was in this case \emph{Respondus} \cite{respondus2021}. \emph{Respondus} installs a restricted version of a web browser called \emph{Lockdown Browser} that only allows connection to the \emph{Moodle} platform of the academic institution. The exam was implemented as a regular \emph{Moodle}  quiz and configured to be used with the proctoring tool. When students clicked on the exam resource, the \emph{Lockdown Browser} took control of their laptops, blocked the screen, closed potentially dangerous applications and showed the \emph{Moodle} quiz. The cloud deployment of \emph{Jupyter Hub} coding framework was easily integrated because it was accessible through a URL. This address was configured as an exception in the \emph{Lockdown Browser}, and the link to access the \emph{Jupyter Hub} service was included inside the \emph{Moodle} quiz. When the students accessed their test, they faced several multiple-choice questions. The students had to answer those questions completing the first part of the exam that tested their knowledge of theoretical foundations. The last question of the quiz was not a real question but a link to the \emph{Jupyter Hub} cloud environment. When clicked, a new tab on the restricted browser opened with the \emph{Jupyter Hub} login screen. Students logged in with provided credentials and then accessed the second part of the exam. Time was controlled by the proctoring tool itself because the students were still inside a \emph{Moodle} quiz. When the time was over, \emph{Moodle} closed the quiz, and \emph{Respondus} closed the browser, so the students could not access the platform anymore. Teachers executed a script that ran backups every 15 minutes and a last one once the time expired. The backup tasks directly access the storage created on the \emph{Kubernetes} cluster, so students do not have to worry about anything once the quiz is finished and the proctoring tool is closed. Figure \ref{fig.exam} shows a diagram of the framework.

The result of this experience was a very close approximation to an actual in-class, test-taking scenario for the students. The only configuration required was the installation of the proctoring tool. Theoretical and practical exams were integrated, and students were not requested to perform any additional tasks once the time expired. 

This setup allowed a single teacher to assess an entire class with 30 students with minimum surveillance effort. Proctoring tools provided the mechanisms to minimise the probability of cheating, and \emph{Moodle} with \emph{Jupyter Hub} cloud environments provided a functional coding infrastructure.

\subsection{The feedback}

After testing the three alternative setups A, B and C, the first setup A required four teachers to implement the surveillance of a similar 30 students exam. The second setup B required a high level of expertise configuring laptops and cell phones the days prior to the exam. IT staff reported an increase in the number of issues opened by students to fix webcams or configure video recording applications. Some students also reported problems uploading large videos with their asymmetric Internet connections.

The new setup C proposed in this document corrected all these drawbacks. It provided an environment with almost no configuration from the students' point of view. It was easy to survey with a minimum number of teachers, and it required no extra intervention from the students once that the time expired. No intervention from the IT department was needed either.

\subsection{The expense}

The modules required to replicate this project are a \emph{Kubernetes} cluster with \emph{Jupyter Hub} and an instance of \emph{OpenLDAP} to authenticate the users. These two modules are sufficient to emulate a computer lab. Appendix \ref{replication} describes the process in more detail. Little extra efforts will be needed if this infrastructure is intended to be used for remote exams. Those efforts would be directed to obtain a proper configuration of the authentication services, customize the permissions given to the students and teachers, setting the encryption of the communications or the automation of programming backups of Kubernetes storage. Most of those tasks can be orchestrated through scripting. As mentioned previously, this may require some expertise; however, frequently scripts are simply a sequence of the commands that are executed in the command line.

\begin{figure}[!ht]
\centering
\includegraphics[scale=1]{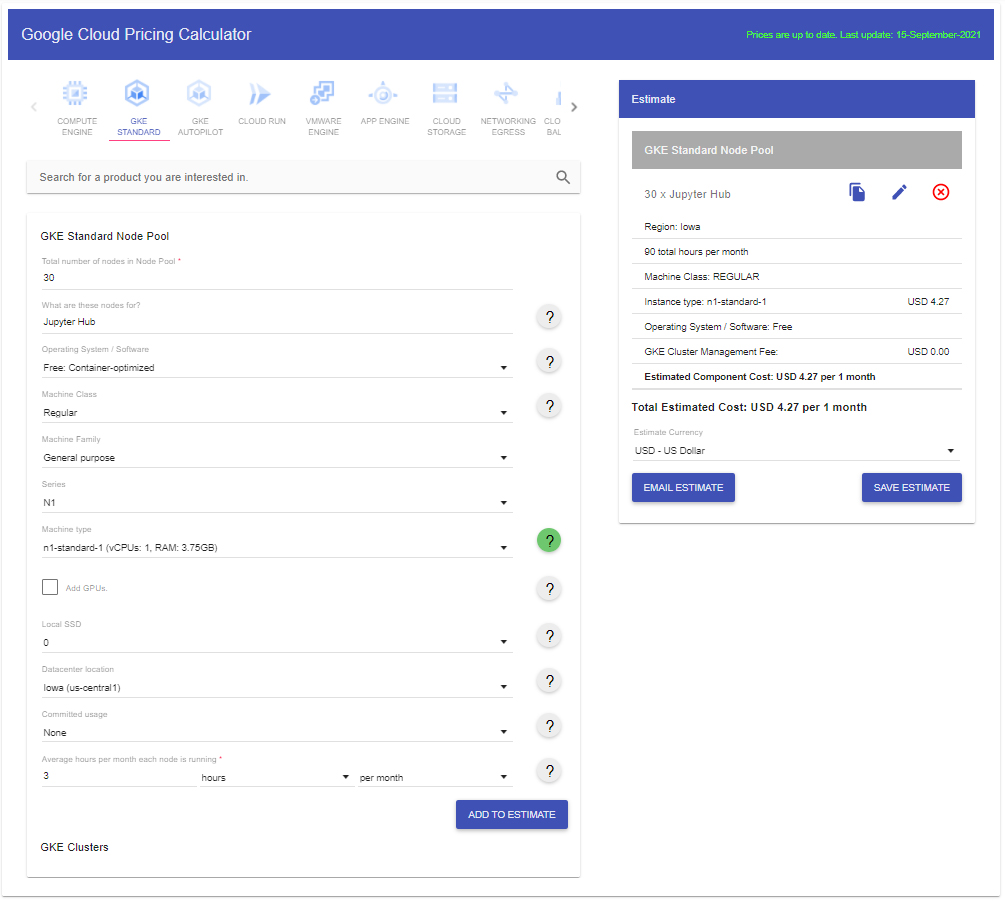}
\caption{Google Kubernetes Engine Pricing}
\label{fig.cluster.pricing}
\end{figure}

The only components required for replicating this project are the \emph{Kubernetes} cluster and an authentication service. The price of such a deployment can be estimated using the cloud provider resources calculator  \cite{calculatorgcp2021}. A conservative approach is shown in Figure \ref{fig.cluster.pricing}. The budgeted cluster consists of 30 general purpose nodes, this deployment allows hosting an exam with the same number of students that the experience described in this document. Default nodes are single CPUs with 3.75GB of RAM. It is a conservative selection of resources because this hardware would be equivalent to a dedicated mid-range laptop for each student. If smaller clusters are desired, better nodes should be chosen with more CPUs and RAM. A standard average university exam's duration is estimated to be two hours. Including an additional hour to set up the environment at the beginning of the exam, capture backups and release the resources at the end, the calculator estimates less than nine dollars. When the cluster is no longer needed, the resources are released, and the institution does not need to support the charges that it assumes with traditional \emph{on-premise} deployments. Table \ref{table.prices} shows alternative deployments with their costs depending on the type of nodes selected. Note that it might be possible to select a single node with 64 CPUs and 256GB of RAM, but the minimum number of nodes for a \emph{Kubernetes} cluster is three.

\begin{table}[ht]
\centering
	\begin{adjustbox}{angle=90,height=\textwidth}
		\begin{tabular}{p{2.5cm}>{\centering\arraybackslash}p{2cm}c>{\centering\arraybackslash}p{2cm}>{\centering\arraybackslash}p{2cm}cp{70mm}}
		    & Number of Nodes  & Node Types   & CPUs per node & RAM per node & Price per 3h & Comments\\
		    High RAM & 4  & n1-highmem-8   & 8 & 52GB & 5,68\$ & More RAM per CPU for intensive computations \\
		    Balanced & 4 & n1-standard-8  & 8 &  30GB & 4,56\$ & One CPU and 3.75GB per student. Few nodes, faster to deploy or update\\ 
		    Conservative & 30 & n1-standard-1 & 1 & 3.75GB & 4,27\$ & Equivalent to a mid-range laptop per student\\
		    Cost optimized & 4 & e2-standard-8 & 8 & 32GB & 3,21\$ & Cheapest nodes without SSD storage\\ 
		    Cheapest & 8 & e2-standard-2 & 2 & 8GB & 1,61\$ & Two students per CPU with cheapest nodes\\
		  \end{tabular}
	\end{adjustbox}
	 \caption{Google Cloud Platform Prices for Different Clusters Deployment Configurations}
	 \label{table.prices}
\end{table}%

\section{Conclusions}
The ultimate objective of this paper is to share the benefits of cloud deployment for online summative assessment within the academic community. Higher education underutilizes these technologies that other sectors of society use extensively. This article is not encouraging teachers to deploy their platform, bypassing the IT departments of their educational institutions. The authors of this document believe that there is still a profound lack of knowledge regarding cloud technologies within academia. Born out of the pressing need to overcome the limitations brought on by the pandemic to evaluate STEM students, cloud deployment resulted in an assessment interface that registered fewer incidents and required fewer human resources than the online exams that used other configurations. 

This assessment platform's simplicity would help increase the frequency of low-stake testing, which is proven to be effective for learning. In addiction exam anxiety is reduced with practice. Those technologies allow the evaluation of practical skills more frequently, allowing for the steady reduction of student anxiety in the face of exams. The cloud deployment described here increases the instructor's ability to easily administer low-stakes or no-stakes testing that would have the added benefit of reducing exam anxiety in the face of final summative assessment. This framework removes the complication of the administration of an exam. The risks of significant interruptions or deterioration to services contribute to an overall reduction in the procedural complexity for the student, resulting in the removal of this specific cause of anxiety.

If current social distancing limitations are finally retracted, these deployments would still be of great interest to academic institutions. Physical space and on-campus facilities could be re-purposed for other institutional needs since computer laboratories would not be necessary for lab assignments or exams. Those tasks could be taken using the student's device, which would be of particular interest to those with visual or hearing impairments. Students with mobility impairments could do both remotely but meeting all the requirements that guarantee an assessment under the same conditions as those who participate on-site. 

In addition, equipment renewal or software updates would not be required of the IT department. This framework could also be adapted to cover other problems, such as reducing compatibility issues between students' computers and the software needed for classes. The number of licenses for specific expensive software could also be decreased if installed in the cloud because students rarely use that application simultaneously. Finally, cloud deployments offer additional opportunities for use with formative assessment and online collaborative learning. Cloud deployment allows for these various use cases, creating an easy-to-use interface that permits the rapid configuration of multiple environments required for an academic institution's resilience and growth in the face of unforeseen developments.

\section*{Abbreviations}
\begin{acronym}
  \acro{API}{Application Programming Interface}
  \acro{CPU}{Central Processing Unit} 
  \acro{GCP}{Google Cloud Platform}
  \acro{GPU}{Graphics Processing Unit} 
  \acro{GUI}{Graphic User Interface} 
  \acro{HTML}{HyperText Markup Language} 
  \acro{HTTP}{Hypertext Transfer Protocol} 
  \acro{IDE}{Integrated Development Environment} 
  \acro{IT}{Information Technology} 
  \acro{IP}{Internet Protocol} 
  \acro{LDAP}{Lightweight Directory Access Protocol} 
  \acro{RAM}{Random Access Memory} 
  \acro{STEM}{Science, Technology, Engineering, and Mathematics} 
  \acro{URL}{Uniform Resource Locator} 
  \acro{VPN}{Virtual Private Network} 
\end{acronym}

\section*{Acknowledgments}

This project is part of the innovation in education tasks carried out by the authors as lecturers at Comillas Pontifical University. The authors are thankful for the support to carry out this research. The content of this manuscript is the solely responsibility of the authors and does not necessarily reflect the position of the funding institution.

\section*{Funding}
This project is part of the innovation in education tasks carried out by the authors as lecturers at Comillas Pontifical University. The authors are thankful for the support to carry out this research.

\section*{Declaration of Competing Interest}
The authors declare that they have no known competing financial interests or personal relationships that could have appeared to influence the work reported in this paper

\section*{Submission declaration and verification}
The authors declare that the work described in this document is original and has not been published previously or is under consideration for publication elsewhere.
The authors also state that if this work is accepted, it will not be published elsewhere in the same form, in English or in any other language, including electronically
 without the written consent of the copyright holder

\section*{Availability of data and materials}
The scripts referenced in the document can be found in the repository associated to this document\footnote{Repository: \url{https://github.com/lmerchante/LowCostUniversalAccess}}.

\appendix
\setcounter{figure}{0}
\section{Tips For Replication}\label{replication}

The purpose of this article is to share this experience with the academic community to enlighten educators who have found themselves in similar circumstances when evaluating their students on coding skills. This document does not intend to train anyone on cloud technologies but to transmit that their use is feasible for everyone with few IT skills. This section provides some references and additional material that educators interested in replicating this experience may find helpful. Extended instructions and the scripts referenced in the document can be found in the repository associated to this document\footnote{Repository: \url{https://github.com/lmerchante/LowCostUniversalAccess}}.

\subsection{Deploying a Kubernetes cluster}

Deploying a distributed infrastructure is not an easy task. Configuring every node of a cluster implies dealing with operative systems, installing the middleware that supports the distributed application, setting the nodes to work on cluster mode, using the proper networking equipment, defining its communication rules and assessing that the platform is secured means a tremendous amount of work.  

Cloud deployments simplify the process to a mere form fulfilling procedure. Some knowledge about the applications to be deployed is still required. But those forms are designed so that almost anyone with a minimal idea can deploy a complex infrastructure like a distributed computation engine.

\begin{figure}[!ht]
\centering
\includegraphics[scale=0.7]{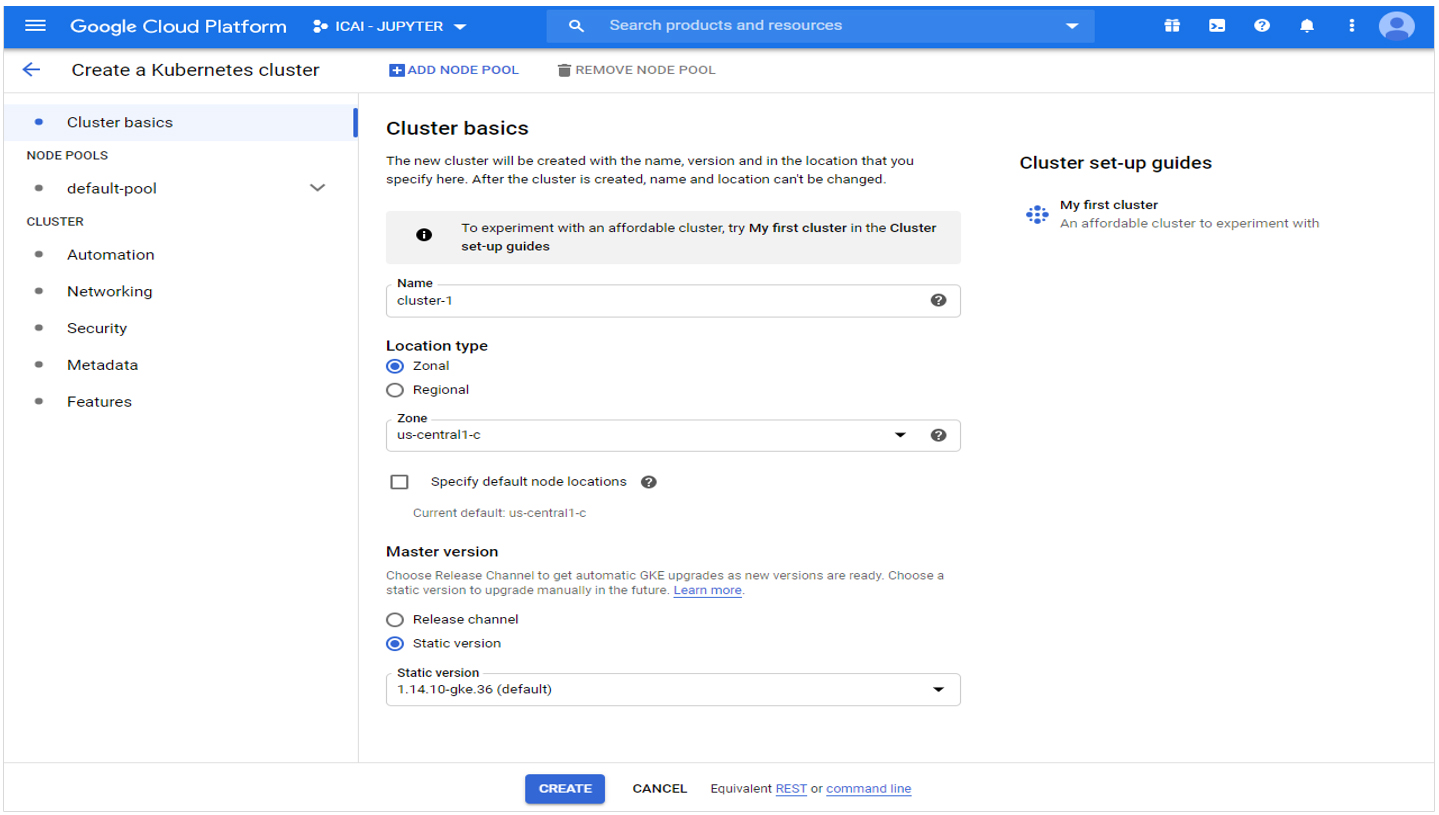}
\caption{Cluster Basics at Kubernetes Cluster Creation Form from Google Cloud Platform}
\label{fig.cluster.creation}
\end{figure}

In this experience, \emph{Google Cloud Platform} was used (GCP from now on). Google provides some free credits for academic members. All cloud providers have this kind of program. The reason to chose GCP was a more significant expiration time for those credits. The process was partially inspired by \cite{kubegcp2021}. The creation is proposed to be accomplished using GCP command line. GCP command line can be used directly on their website or installed locally on a laptop. As explained in the previous section, the main benefit of creating a cluster using commands is that it can be scripted. This mode is recommended because that will allow creating new clusters by running that script. For inexperienced users, it will probably be easier to start using the graphical interface. It can be done at the GCP Portal \cite{portalgcp2021} in the \emph{Compute} section clicking at \emph{Kubernetes Engine}. GCP will offer the possibility of creating a new cluster. In the cluster creation form, shown in Figure \ref{fig.cluster.creation}, only three values are requested:

\begin{itemize}
	\item Cluster name. Name to identify the cluster
	\item Zone or Region. This field has to do with high availability requirements. As detailed in GCP documentation \cite{regionsgcp2021}, putting resources in different zones in a region reduces the risk of an infrastructure outage affecting all resources simultaneously
	\item Versions. GCP offers the possibility of selecting the version of Kubernetes to deploy. Version number might be relevant when connecting the deployment to another infrastructure. But this is not the case
\end{itemize}

\begin{figure}[!ht]
\centering
\includegraphics[scale=0.80]{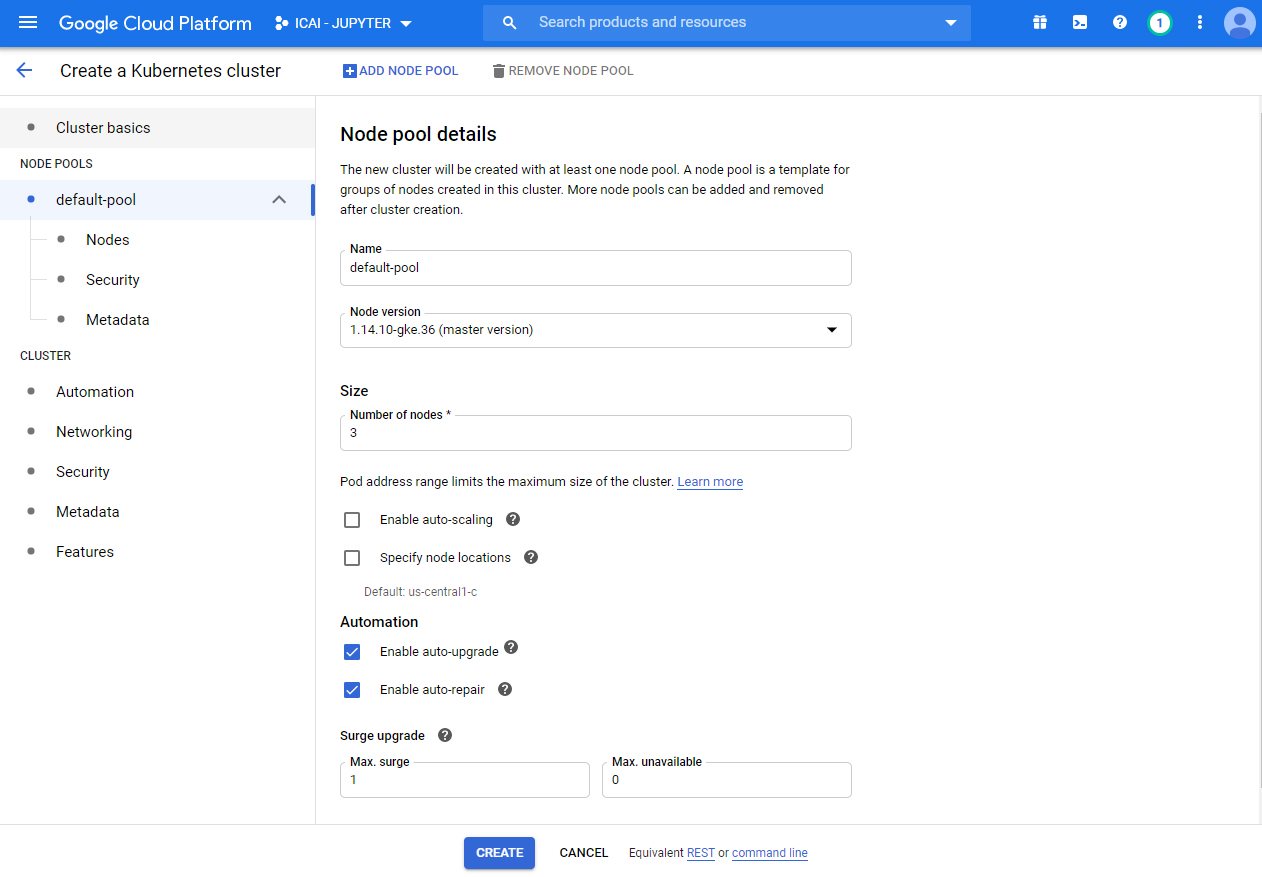}
\caption{Node Pools at Kubernetes Cluster Creation Form from Google Cloud Platform}
\label{fig.cluster.nodepools}
\end{figure}

Those are the minimum entries for \emph{Kubernetes} cluster creation. After giving a name to the cluster, the \emph{Create}  button activates, and the deployment process may start. Extra configuration features can be selected before creating the cluster. On the left-hand side of the GUI there is a navigation panel. The \emph{Node Pools}  section, shown in Figure \ref{fig.cluster.nodepools}, allows configuring the size of the cluster. Again, few parameters are requested:

\begin{itemize}
	\item Pool Name. Name to identify the node pool
	\item Node Version. This version should be the same that the version selected in the Cluster Basics form
	\item Number of nodes. The size of the cluster should be proportional to the number of students that will be serviced. A good starting point is three nodes. This number can be modified later
\end{itemize}

There are a few exciting check-boxes on this form. The first one is \emph{Auto-scaling}. Checking this box means that if the availability of the resources starts to be compromised, the cluster will add new nodes to the cluster without human intervention. Selecting this option may improve performance, but on the other side, adding more hardware resources will increase the bill. This risk can be controlled by choosing the minimum number and the maximum number of nodes. \emph{Auto-scaling} is a desirable feature when there is no intuition about the number of required CPUs or memory that the framework will need or how many students will connect simultaneously. Experienced users can leave this box unchecked and add or remove nodes at will manually. The other interesting check-boxes are the \emph{Enable auto-upgrade} and \emph{Enable auto-repair}. \emph{Enable auto-upgrade} allows the cluster to upgrade the Kubernetes version automatically. \emph{Enable auto-repair} enable the cluster to recover from certain errors without intervention. The correct use of those three check-boxes reduces the maintenance tasks performed by the IT staff to the minimum.

\begin{figure}[!ht]
\centering
\includegraphics[scale=0.8]{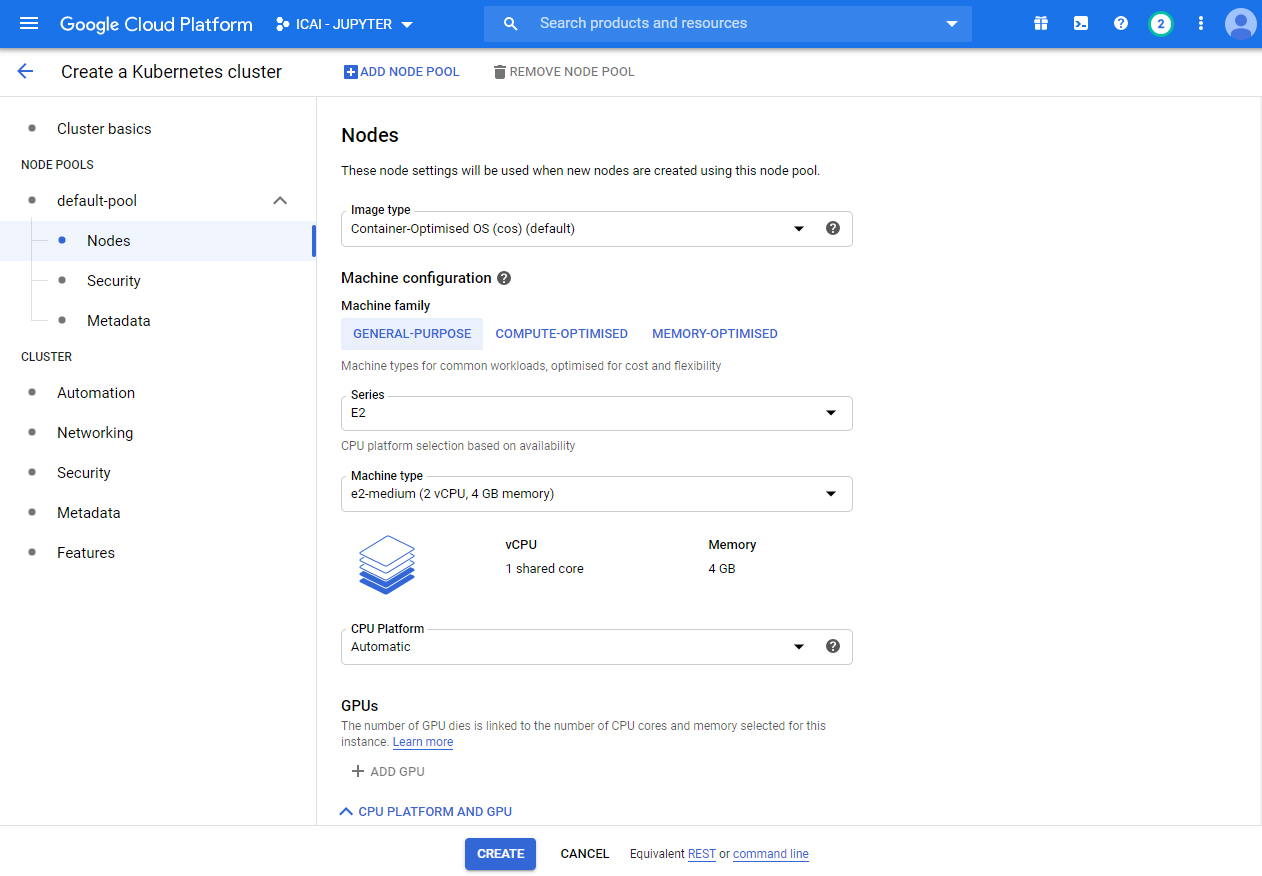}
\caption{Node Types at Kubernetes Cluster Creation Form from Google Cloud Platform}
\label{fig.cluster.nodetypes}
\end{figure}

When configuring the node pool, selecting the number of nodes is critical, but it is not the only relevant parameter. There are different types of nodes depending on the expected use. Nodes are organised on families called \emph{series}, depending on their technology. The default series are built for general purposes with a good combination of CPUs and memory. There are other node series optimised for exigent computation, prioritising the number of CPUs, and some others optimised for memory consuming applications with larger values of RAM. Each series contains different nodes with a great variety of CPUs and RAM values. The type of nodes can be selected with the form shown in Figure \ref{fig.cluster.nodetypes}. Prices vary between series and between nodes inside the same series. Thus, creating the proper cluster with the best resources and price trade-off is, most of the time, an iterative process.

There are many other options for configuring a \emph{Kubernetes} cluster, but there is no need to go deeper to replicate the experience described in this article. After selecting the number and the type of nodes and pushing the create button, it takes around five minutes to get a cloud \emph{Kubernetes} cluster up and running.

\subsection{Installing Jupyter Hub}

The primary reference for installing \emph{Jupyter Hub} on the top of a \emph{Kubernetes} cluster can be found on the \emph{Jupyter Zero} project website \cite{zero2021}. A section called \emph{Setup JupyterHub} has instructions to get this development environment adequately installed on the infrastructure. These instructions are structured on three blocks:

\begin{itemize}
	\item Setting up Helm. Helm is the package manager of Kubernetes \cite{helm2021}. If the cluster was created using GCP forms, Helm had already been installed, and this step can be ignored
	\item Setting up \emph{Jupyter Hub}. This section describes the exact commands that need to be run to have \emph{Jupyter Hub} installed on \emph{Kubernetes}. Having a basic version running is easy. Adding encryption and authentication or using a customised image of Jupyter Hub requires some extra configuration
	\item Tearing everything down. The cluster must be shut down when the activity is over (exams, labs…). Keeping the cluster running would increase the bill. Unfortunately, cloud services cannot be paused to interrupt the billing because the resources would remain assigned to the project. To stop paying for those resources, they need to be released. This block explains how to dismantle everything to stop paying while the cluster is not being used.
\end{itemize}

\subsection{Installing LDAP Service}
LDAP \cite{ldap2021} is one of many authentication possibilities. LDAP is popular and very flexible to adapt to most of the use cases. Many academic institutions use this kind of authentication to let their students access to their services. For this experience, to avoid impacting production servers, a separate LDAP instance was deployed. An independent LDAP service allows custom groups of users with different permissions that could not have been implemented using the production server. For example, a group for students with access only to their Jupyter Hub instances and a group for teachers who would have permission to impersonate any of the students and access their files.

The cheapest LDAP deployment was to install an LDAP server also in the \emph{Kubernetes} cluster. That installation implies no extra budget because the \emph{Kubernetes} cluster is billed by hours of use, not by the number of installed applications. The installation process is very similar to the installation of \emph{Jupyter Hub}. The chosen service was \emph{OpenLDAP} \cite{openldap2021} because it's an open-source implementation of LDAP, lightweight, and well supported. In addition, there exist charts of this product to be installed with \emph{Helm} on \emph{Kubernetes} \cite{openldaphelm2021}. 

The administration of LDAP can be a little tricky. It is possible to find very nice open-source products like \emph{Apache Directory Studio} \cite{apachedirectory2021} that provides a graphical user interface to administrate the LDAP server intuitively.

\bibliography{mybibfile}

\end{document}